# Just Google It

## Digital Research Practices of Humanities Scholars

by Max Kemman, Martijn Kleppe and Stef Scagliola

### Citation

### Abstract

The transition from analogue to digital archives and the recent explosion of online content offers researchers novel ways of engaging with data. The crucial question for ensuring a balance between the supply and demand-side of data is whether this trend connects to existing scholarly practices and to the average search skills of researchers. To gain insight into this process we conducted a survey among nearly three hundred (N= 288) humanities scholars in the Netherlands and Belgium with the aim of finding answers to the following questions: 1) *To what extent are digital databases and archives used*? 2) *What are the preferences in search functionalities* 3) *Are there differences in search strategies between novices and experts of information retrieval?* Our results show that while scholars actively engage in research online they mainly search for text and images. General search systems such as Google and JSTOR are predominant, while large-scale collections such as Europeana are rarely consulted. Searching with keywords is the dominant search strategy and advanced search options are rarely used. When comparing novice and more experienced searchers, the first tend to have a more narrow selection of search engines, and mostly use keywords. Our overall findings indicate that Google is the key player among available search engines. This dominant use illustrates the paradoxical attitude of scholars toward Google: while provenance and context are deemed key academic requirements, the workings of the Google algorithm remain unclear. We conclude that Google introduces a black box into digital scholarly practices, indicating scholars will become increasingly dependent on such black boxed algorithms. This calls for a reconsideration of the academic principles of provenance and context.

# Just Google It

## Digital Research Practices of Humanities Scholars

by Max Kemman, Martijn Kleppe and Stef Scagliola

## 1. Introduction

Throughout the last decade we have witnessed an explosion of available digital databases and archives, due to many digitisation efforts from institutions. The next step currently being undertaken is opening up these archives in new ways, allowing researchers to study more new sources as well as probing new research questions. However, in the development of digital information environments for scholarly use, it is important to learn about users' behaviours, needs and preferences. In this paper we investigate the current research practices of scholars to learn more about how they handle the increasing amount of data and information on the Internet. Do scholars believe the more data the merrier? The following quote from one of our interviews with a PhD-student in History indicates otherwise:

> *"If there is an easier way, I will do it another way. So, I won't go on the Internet to search. There's so much, there is so much information. You can better call someone who knows than search for it yourself."*

Given the large amounts of digital online sources and the impossibility of always being able to trace their origin, recurrent themes in the discussion on the advantages and disadvantages of digitised archives for academic research are the concepts of provenance and context (Brockman, Neumann, Palmer, & Tidline; Kelton, Fleischmann, & Wallace; Lee; Palmer, Teffeau, & Pirmann). *Provenance* refers to the archival principle that data transferred to archival custody retains its distinct character and order attributed to it by its creator (Fickers). *Context* refers to the relation of the data to other entities. A scholar should be able to trace the history of how a document has been documented over time, in order to understand its relation to the organisational, functional, and operational circumstances in which it was created, received, stored and utilised (Pearce-Moses; Kleppe). While both principles are crucial for the interpretation of sources, they can be regarded to be equally relevant for insight into the selection mechanisms of sources, regardless of whether a researcher searches for data to process or literature to gain knowledge on a topic. Reliance on a mechanism to which the intensive work of selection has been delegated implies the risk of de-contextualisation. Algorithms offer such a mechanism. However, many researchers are not aware of the fact that search algorithms are not neutral, but contain the “dispositions, the habitus, the assumptions of its coder” (Hillis, Petit, & Jarrett 5). As such, search engines do not simply retrieve information, but *co-produce* information by ranking and indicating the importance of information. Van Dijck warns that “unawareness of the implications of convenient yet black-boxed tools inevitably leads to more control by owners of search technologies over the production of knowledge” (Van Dijck 587). Due to the black box nature of search systems, scholars can have difficulty understanding the relation between their query and the retrieved results (Buchanan & Cunningham). As also discussed by Hillis, Petit & Jarrett, the only explanation for search results lies in the entered keywords, which de-contextualise the retrieved information in such a way that the search results are only comprehensible through the ordering offered by the search engine.

Of course these issues were equally problematic in the analogue era. The archivist preferences or biases were rarely made explicit in the inventory, yet the scale of what was available was less overwhelming in those days. With the digitisation of collections, not only the scale increased drastically, but the scholar also adopted the selection task from the archivist. The dominant role of search and technology in search environments seems to ask for new proficiencies (Fickers; Vajcner).

*Fragment 1: A Dutch scholar describes what students should learn about performing digital research.*

http://www.youtube.com/watch?v=wKGbUeXrwco

To understand how researchers cope with the growing online information and to further investigate their attitude towards provenance and de-contextualisation within digital search environments, we have investigated the current research practices of Humanities scholars. The goal was to answer the following research questions 1) *To what extent are digital databases and archives used*? 2) *Which search techniques are applied?* And 3) *Do we see differences in search behaviour between experts and novices?*

A survey was held amongst 288 scholars from the Netherlands and Belgium and three additional qualitative interviews were conducted and video-recorded, with the goal of enhancing this paper with short clips containing key statements.

## 2. Related Work

Previous research has shown that the first consequence of the digitisation of collections, archives and search environments, is that scholars are indeed performing their research practices increasingly online. In a survey in the UK (Housewright & Schonfeld, 2013a), it was found that for known-item searches, only about one out of twenty humanities respondents started their research at the library building, whereas almost four out of ten started at a general purpose search engine (such as Google). The remaining respondents started at specific electronic databases (about one out of four), the online library catalogue (about one out of five), or “national or international catalogues or databases” (about one out of seven). For exploratory searches, these figures are slightly different: about half of humanities and social sciences respondents indicated they start at a specific academic database or search engine, with about a fifth indicating they start at a general purpose search engine. These figures are different in the US (Housewright & Schonfeld, 2013b), where for known-item searches a third of humanities respondents indicated for a similar question that they started at a specific scholarly database or search engine, almost a third at the library website, almost a third at a general purpose search engine and about one out of twenty at the library building. For exploratory search, almost half of all respondents started at a specific scholarly database or search engine, while only very small percentages asked colleagues or librarians. Several studies indicate that the main reason for researchers to increasingly rely on the web and use digital resources and search environments as opposed to physical, are the advantages of ease and speed (Bader, Fritz, & Gloning; Bulger et al.; Gibbs & Owens).

*Fragment 2: A Dutch scholar describes the advantages of an online database of Dutch pamphlets in terms of speed and convenience.*

As the database was available online, he had direct access to his material and no longer had to travel to a library in another city.

http://www.youtube.com/watch?v=vPqcwTnqOj0

The other big advantage is increase in scale of coverage. For example, Varshney shows that with the introduction of Google Scholar PhD-students use more literature than before. Furthermore, Google has a central role within the scholarly practices as one of the most used search engines, but very often it is used to find other search tools (Rutner & Schonfeld) or to identify keywords (Gibbs & Owens). This indicates that Google might not be the main research environment, although as a starting point it does play a dominant role in the research and discovery process. At the same time there is little evidence that scholars “take full advantage of the possibilities of more advanced tools” (Bulger et al.). Building upon the insights of the above mentioned studies, this paper attempts to complement the broad picture of how digital archives are searched, by presenting findings amongst Dutch and Flemish Humanist scholars.

## 3. Method

These findings are based on an online survey that was conducted from June to July 2012. Scholars were identified by scrutinising the humanities curriculum of all universities in the Netherlands and of several universities in Belgium. Moreover the editorial boards of the main Dutch humanities academic journals were contacted. All these scholars were invited personally via email (response-rate 24.6%) and at the same time the request to participate was distributed via social media and general mailing lists. In total, 342 respondents participated, including students and information specialists from university libraries. As in this paper the focus is on scholars, this analysis only includes 288 respondents, and leaves out the 54 students and information specialists. For an overview of the demographics of respondents, see Figure 1.

*Figure 1: Demographics of respondents (N=288).*

For the value *discipline*, respondents could choose multiple answers; we report the percentages related to number of respondents giving a certain answer. As respondents gave on average 1.35 answers to this question, the total percentage adds up to 135.1%.

| Gender | | Age | | Country | | Position | | Discipline | |
|---|---|---|---|---|---|---|---|---|---|
| Male | 57.3 | 18-24 | 1.4% | Netherlands | 84.0% | PhD-student | 21.5% | History | 45.1% |
| Female | 42.7 | 25-34 | 34.7% | Belgium | 12.2% | Junior researcher | 2.4% | Social Studies | 16.3% |
| | | 35-44 | 24.7% | Other | 3.8% | Senior researcher | 12.2% | Mass Communication & Documentation | 14.2% |
| | | 45-54 | 18.8% | | | Lecturer | 7.3% | Linguistics and related | 9.7% |
| | | 55+ | 20.5% | | | Postdoc | 9.0% | Literature | 9.7% |
| | | | | | | Assistant professor | 18.4% | Philosophical studies | 6.2% |
| | | | | | | Associate professor | 5.6% | Other | 33.9% |
| | | | | | | Professor | 12.8% | | |
| | | | | | | Other | 10.8% | | |

## 3.1. Information Retrieval Self-efficacy

Generalising the results over all the respondents can yield a distorted picture of their search behaviour. To get a better picture of the variety of search behaviour, we distinguish subgroups on the basis of search expertise (Kemman, Kleppe, & Beunders). Two user characteristics are of interest as they are important determinants of search behaviour: domain knowledge and information retrieval expertise (Wang). As this survey focuses on general search behaviour, we investigate the latter. We do so be measuring information retrieval self-efficacy. Self-efficacy is defined as:

> *“People’s judgments of their capabilities to organize and execute courses of action required to attain designated types of performances. It is concerned not with the skills one has but with judgments of what one can do with whatever skills one possesses”* (Bandura, 1986 391).

This means we do not measure whether a respondent can or cannot perform an action, but whether the respondent is confident that he or she will able to perform an action. Information retrieval self-efficacy thus refers to the judgment of one’s capability to search for information online. In previous research, Internet self-efficacy was found to predict students’ information searching strategies and learning in web-based learning tasks (Tsai & Tsai). Computer self-efficacy was found to influence individuals’ expectations of success, emotional reactions to computers and their actual computer use (Compeau & Higgins). In order to measure one's self-efficacy regarding an activity domain, surveys are developed on which respondent's rate their beliefs in their capabilities (Bandura 2006). To measure the information retrieval self-efficacy of respondents, we developed a survey containing eight 5-point Likert scale questions, ranging from very little confidence to very much confidence. To evaluate this design, we used two datasets; 1) the survey discussed in this paper, but with the inclusion of the responses by students and information specialists at academic institutions (342 responses), and 2) an earlier survey conducted amongst journalists (321 responses) (Kemman et al. 2013). This gives a total N of 663. We evaluated this survey following the five guidelines described by Bandura; 1) discriminating questions, 2) correlations, 3) internal consistency, 4) discriminative ability and 5) predictive ability. First, we eliminated one question where more than half of the respondents chose the same answer. Second, we checked whether all items correlated with each other and the final score. Third, we checked for internal consistency, which resulted in an overall Cronbach $\alpha$=.870. For the Academics the result was $\alpha$=.855 and for the Journalists $\alpha$=.884. Fourth, we tested discriminative ability, by testing whether we could find a difference between the self-efficacy of Academics and that of Journalists. This was done by performing a two-tailed *t*-test, which resulted in a significant difference of $p<.05$, thus allowing us to discriminate between the two user groups. The last and fifth step is still on-going research; as this survey was developed to measure self-efficacy without measuring actual user behaviour, we do not yet have data to test the predictive value of our survey. This evaluation led to the *information retrieval self-efficacy survey* with the following seven questions, which were rated on a 5-point Likert scale ranging from "*very little*" to "*very much*".

I'm confident that I know how to:

1. Use filters to refine search results.
2. Know which search engine would suit my search task best.
3. Appropriately use advanced search options.
4. Learn new functionality without a user guide.
5. Learn new functionality with a user guide.
6. Use Boolean operators.
7. Use Google's search operators.

# 4. Results[1]

## 4.1. Usage of Online Databases and Search Engines

To answer the first research question, we first asked scholars what types of digital data they use. Participants could rate their usage of specific types of data on a 6-point Likert scale, in which "*I don't know it*" is rated lower than "*never*", ranging up to "*very much*". We assume that when the mean score is "*regular*" or higher it is part of the common research practice. We found that only scholarly publications, regular text (e.g. news articles, stories) and still images such as photographs are used regularly or more often by a majority of the respondents. Other types of data such as digitised objects (i.e. not digital-born sources such as museum objects), statistical data and multimedia are used less often. See Figure 2 for a comparison of the mean and mode of responses regarding the use of data types.

*Figure 2: Mean and mode responses to "Which of the following digital data or sources do you use professionally (i.e. for research or lecturing purposes)?". Ordered by mean score (N=288).*

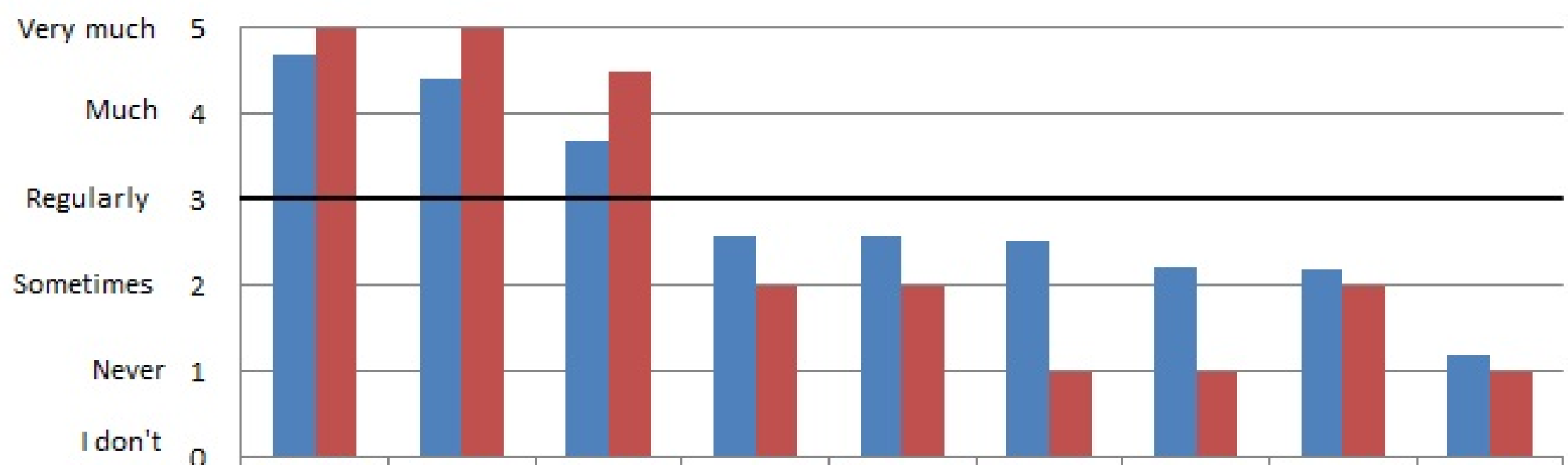

| | Scholarly publications | Text | Images | Video | Statistical data | Digitized objects | Numerical data | Audio | Sensory data |
|---|---|---|---|---|---|---|---|---|---|
| Mean | 4,68 | 4,40 | 3,70 | 2,59 | 2,57 | 2,51 | 2,21 | 2,18 | 1,19 |
| Mode | 5 | 5 | 4,5 | 2 | 2 | 1 | 1 | 2 | 1 |

To learn more about where scholars search these data, we asked respondents to rate on a similar 6-point Likert scale a total of 24 search engines and databases that focus on text, images, scholarly publications and multimedia. As can be seen from Figure 3, in which we show a subset with the more interesting findings, one search company particularly stands out: Google. The first four search engines represent the types of data used by scholars: text, images, scholarly publications and audio-visual content. The only other search system that averaged above “*regular*” is JSTOR,[2] another general search portal; this finding replicates the results by Gibbs & Owens.

*Figure 3: Mean and mode responses to a subset (17/24) of “Which of the following search engines, websites or databases do you use?”. Ordered by mean score (N=288).*

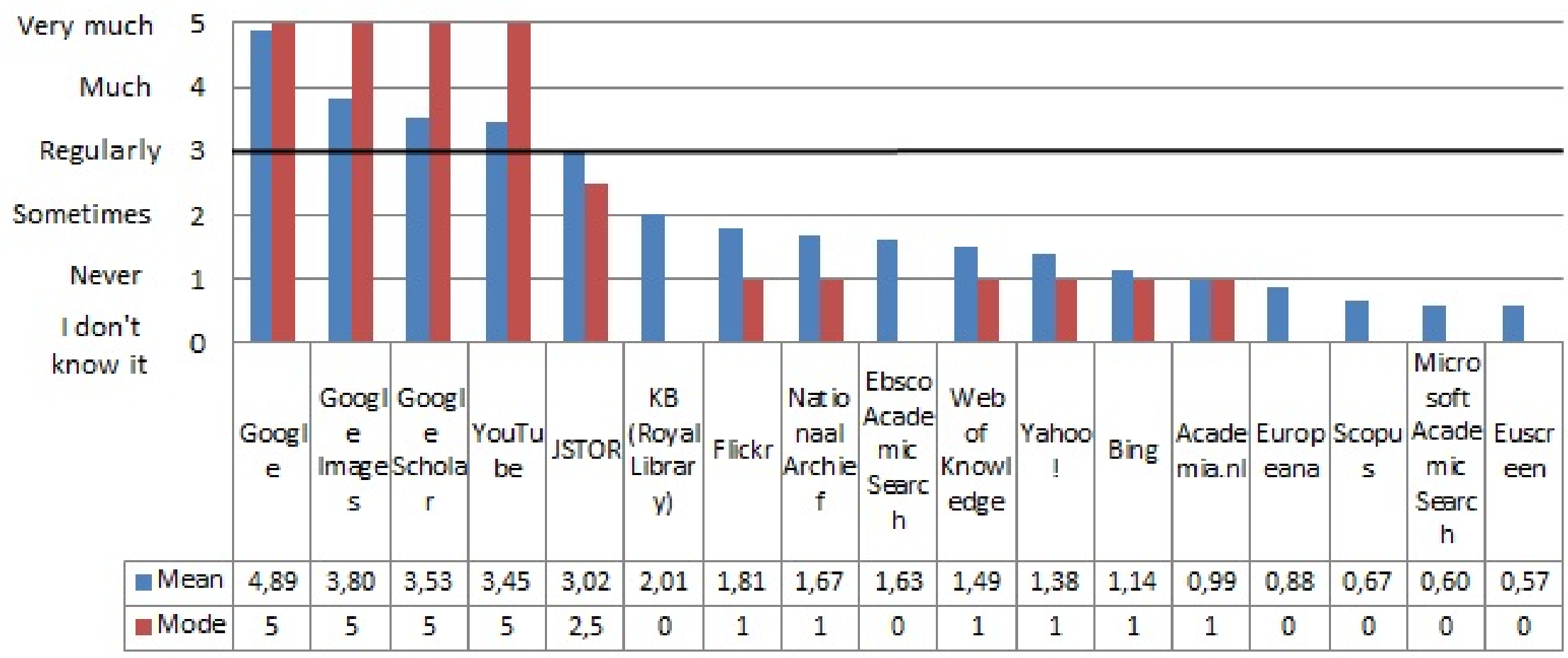

Other findings that stand out concern the largest archives in the Netherlands: the Royal Library (KB)[3] and the Dutch National Archive (Nationaal Archief).[4] On average these archives were used “*sometimes*”, although for KB the mode was “*I don’t know it*”. Other general search engines such as Bing,[5] Yahoo![6] and Microsoft Academic Search[7] score very low and are seldom used. What is striking is that European projects to unlock multimedia archives for research such as Europeana[8] and EUscreen[9] are amongst the lowest scores in use.

Since provenance and contextualisation of sources are crucial concepts in the academic appraisal of knowledge, we evaluated scholars’ reasons to trust search engines and databases. As can be seen in the results shown in Figure 4, previous experience through trial-and-error is the most important reason for trusting search engines, with 79% of respondents indicating this as a consideration for trust. The second most important basis for trust is authority; knowing there is expertise behind the search engine or database. Interestingly, we see that the argument of the search engine offering ‘a broad range of search results’ is not that important, only 25% of respondents chose this value.

*Figure 4: Responses to "When do you trust a search engine or database?". Ordered by frequency (N=288).*

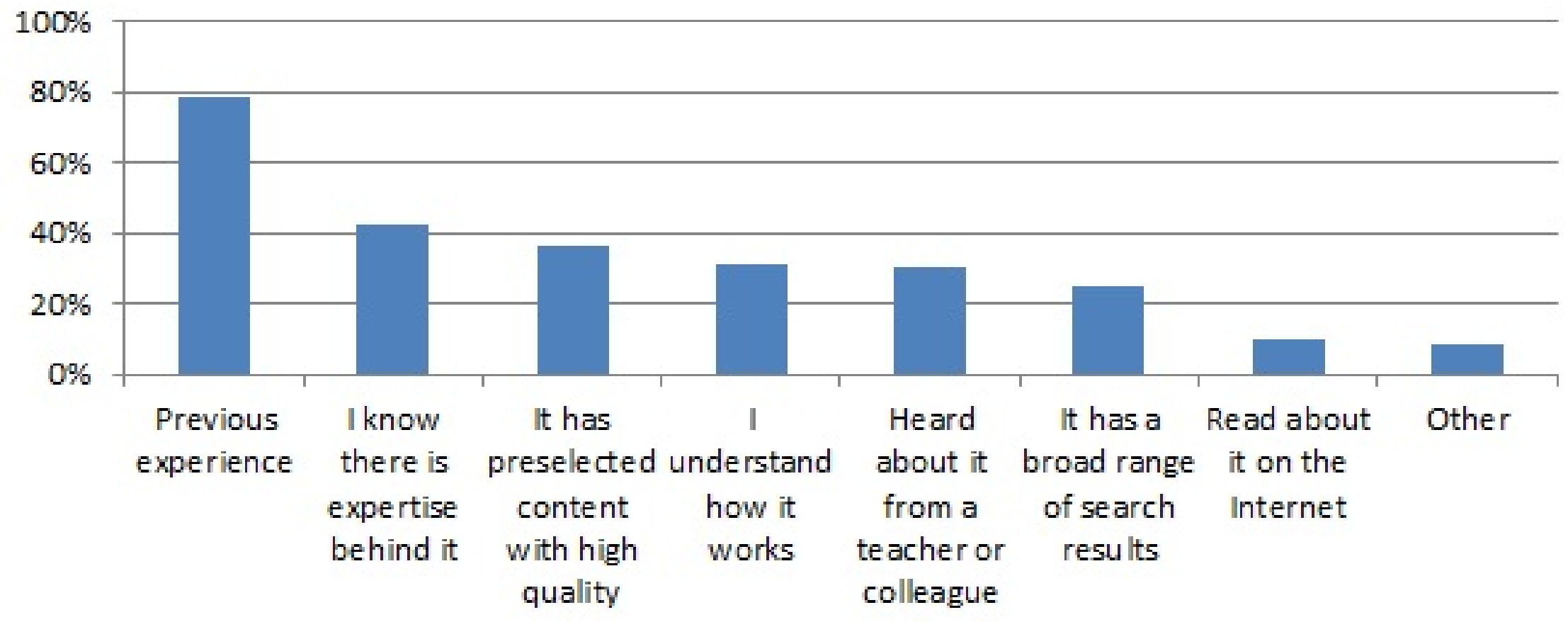

To further analyse the amount of trust respondents have in different search engines and databases, we asked respondents to rate their trust in each search engine and database with a similar 6-point Likert scale as before. We removed the respondents indicating they don't know a search engine or database, leaving a 5-point Likert scale ranging from "*very little*" to "*very much*". In contrast with our earlier finding that Google was used most often, we find respondents do not consider Google as the most trustworthy. JSTOR scores highest, followed by the Dutch archives of the Royal Library and the National Archive. For an overview, see Figure 5 in which we show the search engines and databases from Figure 3 with sufficient respondents. Although the differences are very small, this figure shows that trust due to previous experience cannot fully explain this graph; it appears trust in the authority of an institution and the quality of a collection also play an important role in the trust in a search engine or database. Moreover, trust does not fully explain the usage of a search engine or database; it appears that scholars can still choose to use a search engine even though an alternative is available that they trust more.

*Figure 5: Mean and mode responses to a subset (12/24) of "How much do you trust the following search engines" ordered by mean score.*

Respondents who did not know a search engine or database were removed per case; each bar thus has a different N.

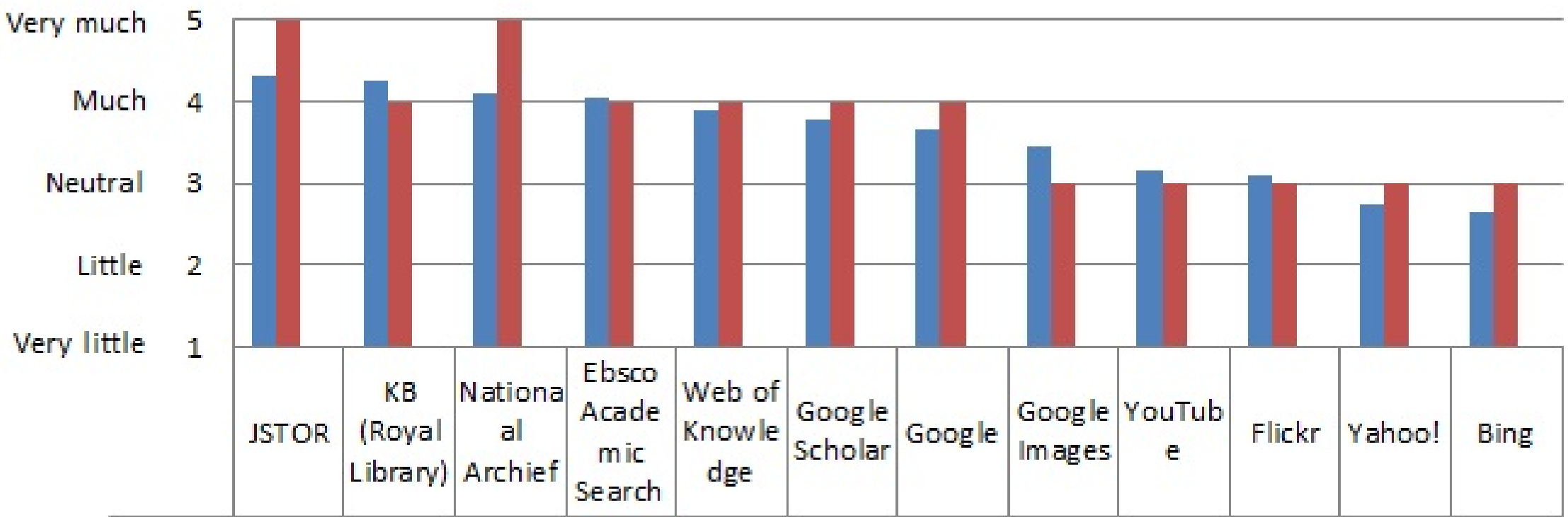

| | JSTOR | KB (Royal Library) | Nationaal Archief | Ebsco Academic Search | Web of Knowledge | Google Scholar | Google | Google Images | YouTube | Flickr | Yahoo! | Bing |
|---|---|---|---|---|---|---|---|---|---|---|---|---|
| Mean | 4,32 | 4,24 | 4,09 | 4,05 | 3,90 | 3,78 | 3,67 | 3,45 | 3,15 | 3,09 | 2,73 | 2,66 |
| Mode | 5 | 4 | 5 | 4 | 4 | 4 | 4 | 3 | 3 | 3 | 3 | 3 |
| N | 243 | 171 | 169 | 127 | 124 | 258 | 279 | 261 | 259 | 184 | 199 | 154 |

## 4.2. Search Functionality

The second sub-question concerns how these search engines and databases are searched, and what search functionalities are used. We asked respondents to rate a number of search functionalities on a similar 6-point Likert scale as before, in which "*I don't know it*" is rated lower than "*never*", ranging up to "*very often*". We again assume that when the mean is "*regularly*" or higher, this search functionality can be assumed to be part of the common search process. This resulted in Figure 6 below, in which we find that only keywords and advanced search options (i.e. a separate page for more advanced search queries) can be assumed to be part of the common research process.

*Figure 6: Mean and mode responses to "While searching the web, which of the following options do you use?". Ordered by mean score (N=288).*

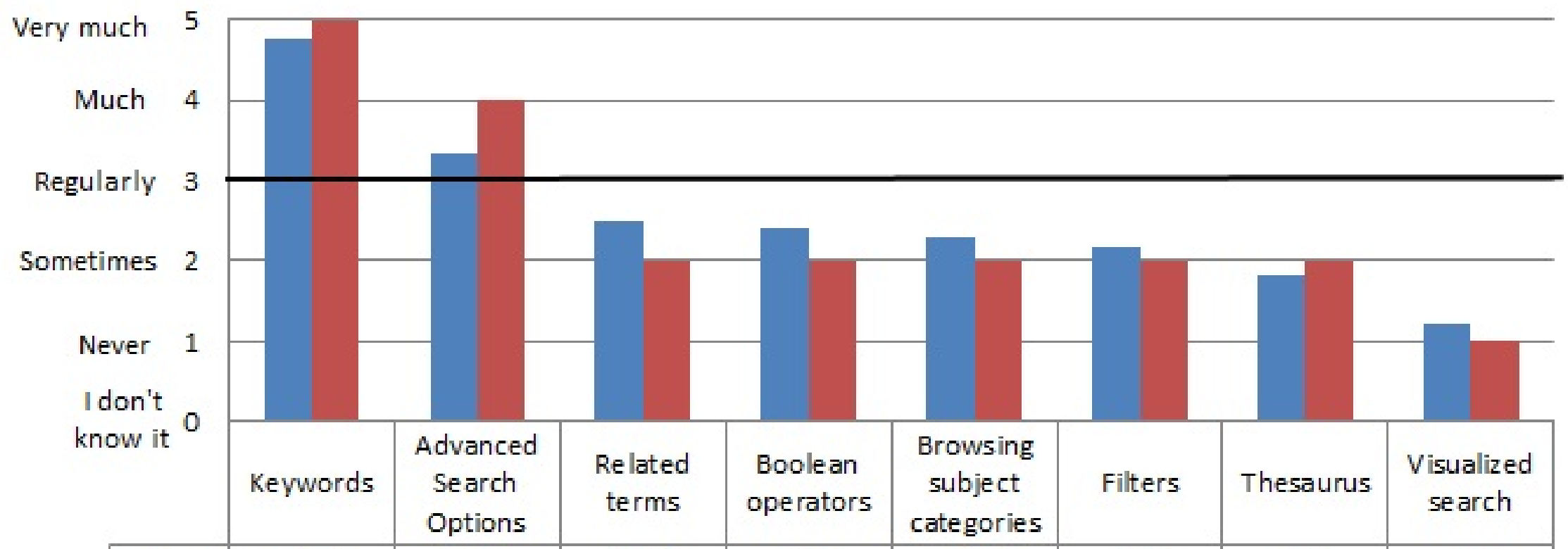

| | Keywords | Advanced Search Options | Related terms | Boolean operators | Browsing subject categories | Filters | Thesaurus | Visualized search |
|---|---|---|---|---|---|---|---|---|
| Mean | 4,77 | 3,35 | 2,50 | 2,40 | 2,28 | 2,17 | 1,83 | 1,21 |
| Mode | 5 | 4 | 2 | 2 | 2 | 2 | 2 | 1 |

Most respondents indicated to only use Boolean operators sometimes. Filters (also known as facets) are only used sometimes, contrary to what has been concluded in previous studies (e.g. Kules et al.). Finally, the browsing of subject categories, more indicative of exploratory search rather than of known-item search, also appears to be performed much less often than keyword search. This also contradicts findings of previous research on scholarly search behaviour (Stevens; Stone).

## 4.3. Differences Between Experts and Novices

Up until now, we have treated all 288 respondents as a single group. However, we are interested to know whether there are differences between scholars with a higher information retrieval expertise and scholars with a lower information retrieval expertise. To get an idea of which respondents score higher in information retrieval self-efficacy, we performed ANOVAs for age and position as shown in Figure 1. From these analyses, we found there is a significant difference with large effect between the age-groups with $F(4, 283)=12.412$, $p<0.01$, $\eta^2=0.149$. From the post-hoc Bonferroni analysis, we found that the group of respondents of 55+ had significantly lower self-efficacy than all other groups. We also found a significant difference with medium effect between positions with $F(8, 279)=3.292$, $p=0.01$, $\eta^2=0.086$. From the post-hoc Bonferroni analysis, we found that PhD-students scored significantly higher than Professors and Senior Researchers. To analyse the influence of information self-efficacy on search behaviour, we divide the respondents in two groups. Respondents scoring below average are defined as novices, whereas respondents scoring above average are defined as experts, drawing upon the work of Tsai & Tsai. See Figure 7 for the distribution of respondents. For the Academics, we found on a scale of 0-4 a mean score of 2.48 (S.D.=.77). We define respondents scoring below 2.48 as *novices* (N=142, 49.3%), and respondents scoring above 2.48 as *experts* (N=146, 50.7%).

*Figure 7: Distribution of participants for information retrieval self-efficacy. The line indicates the cut-off point at the mean (N=288).*

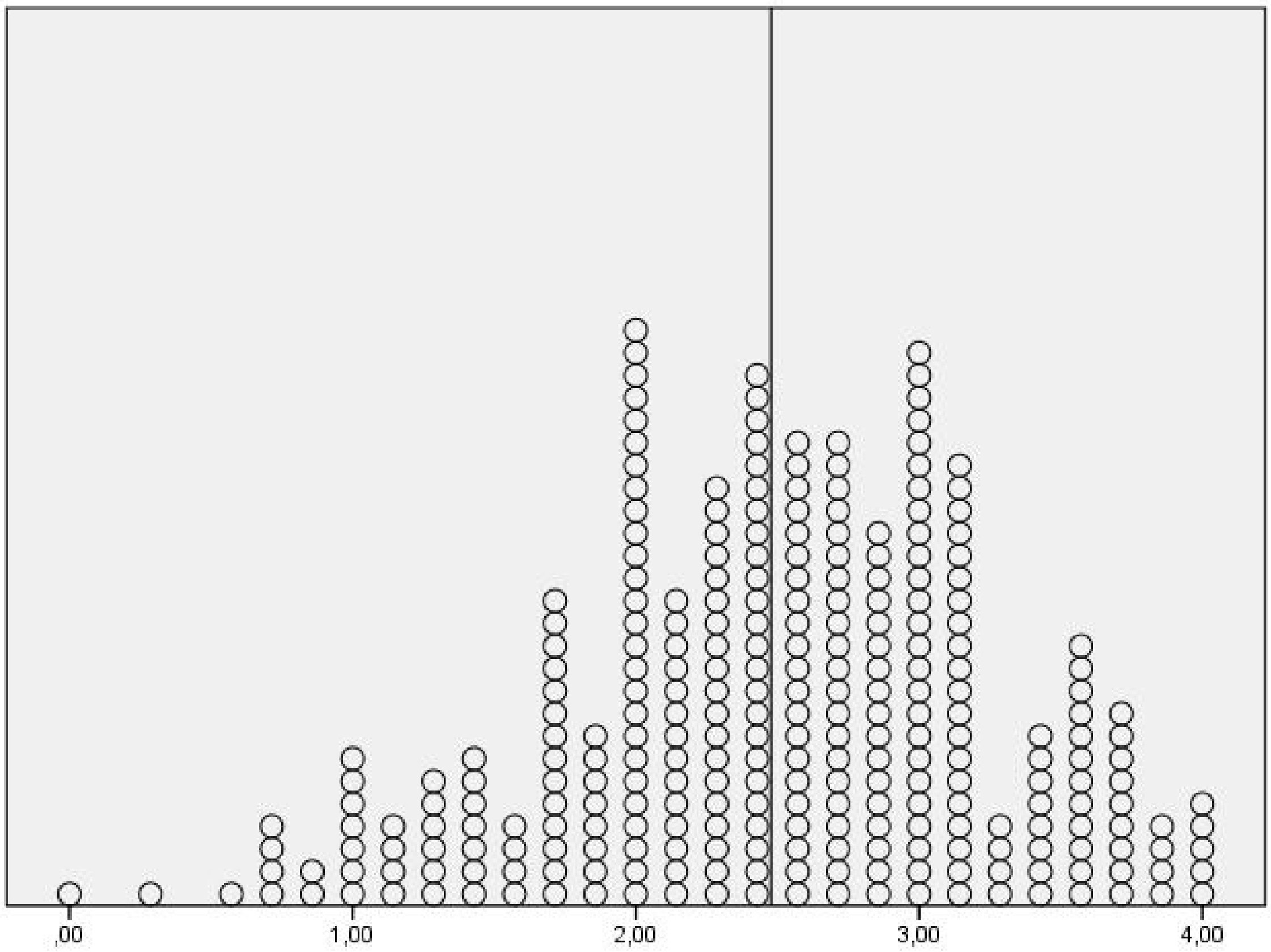

We first consider the difference between the types of digital data used by scholars. We analyse this with a MANOVA with Pillai's trace, with which we find a significant difference with medium-sized effect exists between novices and experts with F(9, 278)=2.067, $p$=.033, $\eta_p^2$=.063. Looking at the tests of between-subjects effects, we see this difference can be traced to the data types numerical data, with F(1)=7.504, $p$=.007 and digitised objects, with F(1)=5.259, $p$=.023.

Considering where scholars search, we performed a MANOVA with Pillai’s trace for the 17 search engines and databases shown in Figure 3. From this analysis, we found a significant difference with medium-sized effect exists between experts and novices with $F(17, 270)=2.202$, $p=.005$, $\eta_p^2=.122$. Looking at the tests of between-subjects effects, we found this difference can be traced to differences for Google Scholar with $F(1)=6.730$, $p=.010$, Flickr with $F(1)=9.047$, $p=.003$, Ebsco Academic Search with $F(1)=11.851$, $p=.001$ and finally Bing with $F(1)=6.772$, $p=.010$. In all cases, experts indicate higher usage than novices, showing a wider usage of search engines and databases. Interestingly, this appears to show that higher expertise in information retrieval does not mean scholars are more critical on the provenance of their sources and more narrow in their choices, but instead they employ a wider choice of search engines and databases.

With regards to the use of search functionality, we performed a MANOVA with Pillai’s trace and again found a significant difference with large-sized effect exists between experts and novices, with $F(8, 279)=11.549$, $p<.001$, $\eta_p^2=.249$. From the tests of between-subjects effects, we found significant differences exist for all the search functionalities except the use of keywords and thesauri.

In short, we found experts are more used to using numerical data and digitised objects, use a wider range of search engines and databases, and use a wider range of search functionalities to perform their online searches.

## 5. Conclusion

We found that text (including scholarly publications) and images are the main type of sources that are searched by scholars, in accordance with Bulger et al. Moreover, in accordance with Gibbs & Owens, we found that general purpose search engines dominate the search practices of scholars, especially Google and JSTOR. Previous experience is the most common reason to trust a search engine or database, although authority and the quality selection of an institution also appear to play an important role. Scholars rely heavily on keywords in their search process, reflecting the dependence on Google-like search engines, although they regularly use advanced search pages when available. Scholars with a higher information retrieval self-efficacy use a wider range of search engines and search functionality, indicating that expertise in information retrieval can be a valuable asset for scholars.

When we combine all our findings, it appears that Google is the key player that enables scholars to search for information online. Scholars want to search for text, images and scholarly articles; Google offers the most commonly used search systems for each. With regard to audio-visual content, which is not yet commonly used in academic research, it is remarkable to note that large European projects to unlock multimedia archives for a broad audience such as Europeana, are used much less than Google's YouTube. This all supports our main conclusion that digital research practices of Humanities scholars in the Netherlands can be condensed to three words: Just Google it.

When reviewing these findings in the light of scholarly practices as laid out in the introduction an interesting paradox emerges. Although provenance and context are deemed key academic qualities, these do not appear to be common considerations in digital research practices. Not only is it difficult to understand what is being indexed by Google, the algorithms that retrieve and rank search results are so complex that in the further development of the algorithm *"it is no longer possible to know exactly how any given change affects the algorithmic matrix as a whole"* (Hillis, Petit, & Jarrett 18). As such, Google introduces a black box into the digital research practices of scholars, but interestingly enough this does not seem to influence the trust of the majority of scholars in search results.

## 6. Discussion

Why do scholars have a strong preference for Google while this may affect the academic principles of provenance and context? In this context the findings by Xie offer insight: in evaluations of information systems, content and usability seem to be the key factors. Due to its domination Google is able to determine what the quality of searching is in terms of content and usability (Hillis, Petit, & Jarrett 52). It is probable that considerations of convenience supersede the principles of provenance and context. Google might not cover all relevant sources, but it does probably cover the most. Furthermore, in terms of efficiency, relying on Google instead of searching in multiple alternative more refined search systems within the websites of specialised institutions, and subsequently comparing the results, saves time and energy. The considerations that we would like to suggest on the basis of our findings with regard to the further development of information systems for the Digital Humanities have a short-term and long-term perspective.

The short-term perspective relates to the usability of Google. In the development of digital Humanities collections and search interfaces, developers should incorporate the notion that scholars assume all collections are findable through Google. In the cases that a collection has its own specific search interface, scholars assume it will work similarly to Google: they expect to see a search box, and expect to be able to use it in standard ways (Halavais). When a database does not meet the expected usability, scholars can fall back on the analogue search environment, as illustrated in Fragment 3.

*Fragment 3: A Dutch scholar describes how he sometimes fell back on a paper catalogue of pamphlets, because the search interface did not meet his expectations.*

http://www.youtube.com/watch?v=L5k5wPgahtI

The long-term perspective that we envision is related to the paradox of using a black box within academic practices: the dependence on Google already shows that academic practices are changing into an age where there is too much data to be able to manually gather and analyse every single relevant record, as illustrated by the quote in the introduction. This may lead to the conclusion that increase in coverage due to technological improvement may lead to altering principles that were formed in the analogue age. This conclusion resonates in the statement expressed by Mozilla's executive director Mark Surman that "current practices are still 'rooted in the analog age'",[10] as agreed by one of our interviewees:

*Fragment 4: A Dutch scholar describes how the digitisation of research practices has not yet changed the way research works.*

http://www.youtube.com/watch?v=-dfBvYIARBw

In the future, we expect an increase in the application of computational analyses on data such as automatic pattern recognition, as has already been realised in many Digital Humanities projects. This calls for a reconsideration of the academic principles of provenance and context, since black-boxed algorithms will be necessary to help scholars find their way in an overwhelming world of information.

### 6.1. Future research

This paper presents the first results of our research on how scholars search and use digital research environments. More research is required in order to gain a deeper insight in how scholars use Google that cannot be acquired through a survey. In the future, we plan to investigate whether the use of Google is related to specific stages of search and research, e.g. is Google mainly used during initial exploration of a topic, or during the complete research process? Moreover, we plan to gain a better understanding of the information need at Google; is it an access point to discover databases of relevance, or does it serve as a one-stop shop of information where material is found directly at Google?

## 7. Acknowledgements

We are grateful for the financial support from the EU FP7 project AXES – *Access To Audiovisual Archives* ICT-269980. We also thank the following colleagues and organisations for their cooperation and input: Henri Beunders, Franciska de Jong, Renske Jongbloed, het Huizinga Instituut, Netherlands Graduate School of Linguistics (LOT) , Nederlands Instituut voor Beeld en Geluid, Onderzoekschool Politieke Geschiedenis, de Vereniging Geschiedenis, Beeld en Geluid.

**Footnotes**

[1] All quantitative survey data are available open access via Kemman, M., Kleppe, M., Scagliola, S. (2013) Just Google It - Digital Research Practices of Humanities Scholars (Dataset). Available at http://www.persistent-identifier.nl/?identifier=urn%3Anbn%3Anl%3Aui%3A13-9x3b-pa

[2] www.jstor.org

[3] www.kb.nl

[4] www.nationaalarchief.nl

[5] www.bing.com

[6] www.yahoo.com

[7] academic.research.microsoft.com

[8] www.europeana.eu

[9] www.euscreen.eu

[10] Mark Surman. Introducing the Mozilla Science Lab. *The Mozilla Blog* 14 June 2013. Retrieved June 22, 2013, from http://blog.mozilla.org/blog/2013/06/14/5992/